\documentclass[journal]{IEEEtran}
\usepackage{graphicx}
\usepackage{float}
\usepackage{amsmath}
\usepackage{algpseudocode}
\usepackage{algorithm}
\usepackage{tabularx}
\usepackage{array}     
\usepackage{booktabs}
\usepackage{romannum}
\usepackage{multirow}
\usepackage{makecell}
\usepackage[switch]{lineno}

\ifCLASSINFOpdf
\else
\fi
\begin{document}

\newcolumntype{C}{>{\centering\arraybackslash}X}
\setlength{\tabcolsep}{12pt}
\renewcommand{\arraystretch}{1.5}

%
\title{Sparsity-Aware Streaming SNN Accelerator with Output-Channel Dataflow for Automatic Modulation Classification}
%
%
%

\author{Kuilian~Yang,~\IEEEmembership{Student Member,~IEEE,}
        Li~Zhang,
        Ahmed~M.~Eltawil,~\IEEEmembership{Senior Member,~IEEE,}
        and~Khaled~Nabil~Salama,~\IEEEmembership{Senior Member,~IEEE}
\thanks{Kuilian Yang, Li Zhang, Ahmed~M.~Eltawil, and Khaled Nabil Salama are with the Division of Computer, Electrical and Mathematical Sciences and Engineering, King Abdullah University of Science and Technology, Thuwal 23955, Saudi Arabia (e-mail: kuilian.yang@kaust.edu.sa; li.zhang@kaust.edu.sa; ahmed.eltawil@kaust.edu.sa; khaled.salama@kaust.edu.sa).}
}

%
%

\markboth{Journal of \LaTeX\ Class Files,~Vol.~14, No.~8, August~2015}%
{Shell \MakeLowercase{\textit{et al.}}: Bare Demo of IEEEtran.cls for IEEE Journals}
%



\maketitle

\begin{abstract}

The rapid advancement of wireless communication technologies, including 5G, emerging 6G networks, and the large-scale deployment of the Internet of Things (IoT), has intensified the need for efficient spectrum utilization. Automatic modulation classification (AMC) plays a vital role in cognitive radio systems by enabling real-time identification of modulation schemes for dynamic spectrum access and interference mitigation. While deep neural networks (DNNs) offer high classification accuracy, their computational and energy demands pose challenges for real-time edge deployment. Spiking neural networks (SNNs), with their event-driven nature, offer inherent energy efficiency, but achieving both high throughput and low power under constrained hardware resources remains challenging. This work proposes a sparsity-aware SNN streaming accelerator optimized for AMC tasks. Unlike traditional systolic arrays that exploit sparsity but suffer from low throughput, or streaming architectures that achieve high throughput but cannot fully utilize input and weight sparsity, our design integrates both advantages. By leveraging the fixed nature of kernels during inference, we apply the gated one-to-all product (GOAP) algorithm to compute only on non-zero input-weight intersections. Extra or empty iterations are precomputed and embedded into the inference dataflow, eliminating dynamic data fetches and enabling fully pipelined, control-free inter-layer execution. Implemented on an FPGA, our sparsity-aware output-channel dataflow streaming (SAOCDS) accelerator achieves 23.5 MS/s (approximately double the baseline throughput) on the RadioML 2016 dataset, while reducing dynamic power and maintaining comparable classification accuracy. These results demonstrate strong potential for real-time, low-power deployment in edge cognitive radio systems.

\end{abstract}


%
\IEEEpeerreviewmaketitle


 




\section{Introduction}
%
%
%
%

 

\IEEEPARstart{T}{he} rapid expansion of Internet of Things (IoT) devices has enhanced daily life but also introduced vulnerabilities to malicious attacks in untrusted electromagnetic environments\cite{hao2023automatic}. Automatic modulation classification (AMC) serves as a critical tool for identifying the modulation formats of potential attack signals, thereby aiding in spectrum risk monitoring and threat detection. AMC's applications span both military and civilian domains, including spectrum management, electronic reconnaissance, and electronic countermeasures~\cite{huynh2021automatic}. In modern communication systems, especially those utilizing software-defined radios, efficiently identifying modulation types is essential for effective signal demodulation and overall system performance. 

Deep learning-based AMC techniques have shown strong performance across various applications~\cite{meng2018automatic, tridgell2019real, wang2020distributed}. However, the high sampling rates of radio signals pose challenges for remote processing due to bandwidth limitations. As a result, deploying AMC directly on embedded edge devices has become a practical and efficient solution for real-time systems. Although traditional artificial neural networks (ANNs) achieve high accuracy, their computational and memory demands make them less suitable for resource-constrained edge platforms~\cite{restuccia2019big}. In contrast, spiking neural networks (SNNs) offer significant potential for low-power operation, thanks to their event-driven nature. In SNNs, the binary input feature map (IFM) allows dense multiply-accumulate (MAC) operations to be replaced with simpler accumulations, reducing computational complexity—particularly important given the hardware cost of multipliers. The event-driven behavior also introduces high temporal sparsity in IFMs. When combined with spatial sparsity in kernels—enabled by training techniques such as pruning and quantization—SNNs provide a promising pathway for developing power-efficient, high-throughput, and real-time AMC systems suitable for edge deployment.

Due to their unique processing patterns, traditional units like CPUs and GPUs often struggle to execute SNNs efficiently. As a result, specialized digital hardware accelerators have emerged as promising solutions, broadly categorized into systolic-array-based and streaming architectures.

Systolic-array-based accelerators employ a fixed grid of homogeneous processing elements (PEs) to execute computations across different SNN layers. They typically rely on centralized control logic—such as routers and schedulers—to distribute inputs and weights, enabling accumulation-based operations. Notable designs like STICKER~\cite{yuan2019sticker}, SNAP~\cite{zhang2020snap}, ESSA~\cite{kuang2022essa}, and LoAS~\cite{fang2024energy} leverage spatial and/or temporal sparsity to skip redundant computations. However, their layer-wise scheduling and frequent memory accesses limit inter-layer parallelism, which constrains throughput.

In contrast, streaming architectures instantiate each SNN layer heterogeneously with locally stored weights, enabling continuous dataflow between layers without external memory access. This eliminates the need for global control logic, thereby reducing latency and improving throughput. Designs such as S2N2~\cite{khodamoradi2021s2n2} and other FINN-based implementations~\cite{guo2023end} exemplify this approach. However, without fine-grained control logic, they often fail to skip zero weights or inputs effectively, and primarily exploit only temporal sparsity—leading to redundant computations that reduce energy efficiency.

To overcome the trade-offs between sparsity exploitation and throughput in existing architectures, we propose a sparsity-aware output-channel dataflow streaming (SAOCDS) accelerator tailored for SNN-based AMC systems. 

To the best of our knowledge, this is the first fully-pipelined streaming SNN accelerator that integrates both temporal and spatial sparsity, achieving automatic workload balancing without resorting to additional control logic, thereby enabling high throughput under resource constraints.

The proposed design achieves a throughput of 23.5 MS/s on the RadioML 2016 dataset~\cite{o2016radio}, while maintaining relatively low dynamic power consumption and competitive classification accuracy.

The key contributions of this work are summarized as follows:
\begin{enumerate}
\item We present a streaming architecture that simultaneously exploits temporal and spatial sparsity, yielding more power-efficient computation while preserving high throughput.
\item We adopt an output-channel dataflow strategy, which inherently balances computational workload and simplifies inter-layer data transmission.
\item We develop a lightweight error-detection algorithm to detect and handle potential logic conflicts, ensuring seamless inter-layer dataflow without incurring additional control overhead.
\item We demonstrate, through implementation on AMC tasks (e.g., RadioML 2016), that our SAOCDS accelerator delivers both high throughput and competitive accuracy under limited hardware resources, illustrating a viable new deployment option for SNNs in resource-constrained, throughput-sensitive scenarios.
\end{enumerate}

The remainder of this paper is organized as follows. In Section~\Romannum{2} we review related work and identify existing opportunities for sparsity utilization. Section~\Romannum{3} analyzes the key challenges of integrating both temporal and spatial sparsity in a streaming architecture, then presents our proposed sparsity-aware output-channel dataflow streaming (SAOCDS) SNN accelerator. Section~\Romannum{4} describes the trained SNN architecture for AMC and reports its classification results. Section~\Romannum{5} details our FPGA implementation and experimental evaluation. Finally, Section~\Romannum{6} concludes the paper.

\section{Background}

In this section, we first survey existing SNN accelerator designs to assess how they exploit sparsity and to identify their limitations. We then highlight remaining opportunities for improving efficiency, which motivate our proposed design.

\subsection{Related SNN Accelerator Work}

\begin{figure}
    \centering
    \includegraphics[width=1\linewidth]{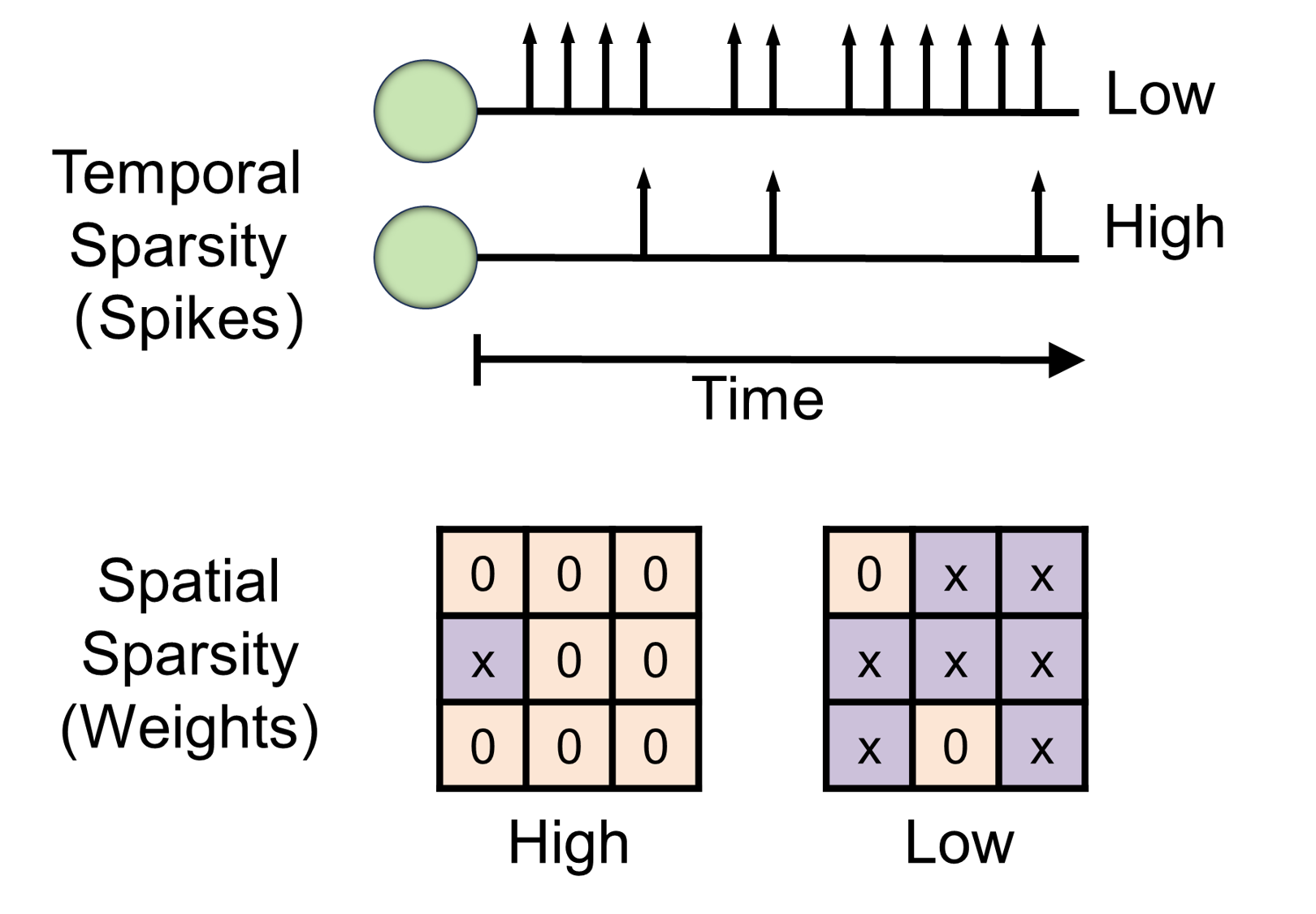}
    \caption{Temporal sparsity and spatial sparsity in SNN}
    \label{fig: sparsity in SNN}
\end{figure}

The inherent sparsity in SNNs manifests itself primarily in two forms (Fig.~\ref{fig: sparsity in SNN}): temporal sparsity, arising from the event-driven nature of spike generation when neurons fire only upon reaching a threshold; and spatial sparsity, referring to zero-valued synaptic weights, often the result of pruning or quantization during training.

Various accelerators have been proposed to exploit sparsity in SNNs. STICKER~\cite{yuan2019sticker} and SNAP~\cite{zhang2020snap} use an input-priority approach, compressing non-zero IFM elements with coordinate metadata to retrieve corresponding weights. However, this increases storage overhead, as 1-bit input spikes require multiple bits of metadata. ESSA~\cite{kuang2022essa} exploits both temporal and spatial sparsity by scanning input rows and using a compressed sparse row (CSR) format for weights. Yet, the inclusion of extensive metadata still introduces considerable overhead. LoAS~\cite{fang2024energy} adopts a two-stage bitmap encoding to simplify IFM storage and reduce redundant fetches. However, since output pixels are computed individually with multiple timesteps processed in parallel, a load balancer is needed to evenly distribute workloads across memory banks.

All of these designs adopt an input-priority strategy, where runtime metadata is used to locate corresponding weights. This leads to latency and logic overhead due to real-time decoding during input streaming. In contrast,~\cite{lien2022sparse} proposes a weight-priority approach based on the gated one-to-all product (GOAP) algorithm, which iterates over non-zero weights and uses precomputed metadata to fetch the relevant inputs. Since weights are fixed during inference, this method reduces runtime computation. However, like prior designs, it remains built on a traditional systolic array, which limits inter-layer parallelism, requires frequent memory accesses, and still depends on control components such as routers and schedulers. Despite their ability to skip redundant operations and improve energy efficiency, these sparsity-aware accelerators fall short of fully leveraging streaming throughput due to their underlying architectures.

To improve inference performance, several SNN accelerators have adopted FINN-based streaming architectures~\cite{guo2023end, khodamoradi2021s2n2}, where each layer is instantiated as a dedicated hardware module with weights stored locally. This enables direct inter-layer dataflow without writing intermediate results to global memory, reducing memory latency and eliminating the need for runtime decoders. Although this approach increases resource usage due to the instantiation of all layers, it achieves high throughput. However, lacking control logic to exploit sparsity, all inputs are processed uniformly—regardless of whether they contribute meaningfully to the output. As a result, the architecture may perform redundant computations, limiting its efficiency despite its high throughput.

In summary, current accelerator designs trade off computational efficiency against throughput: control logic can skip redundant operations but may introduce overhead that degrades speed. Combining inter-layer parallelism to sustain high throughput with sparsity-aware skipping of pointless operations thus remains a key challenge.

\subsection{Opportunities of Sparsity Utilization}

Upon analyzing the operational principles of existing streaming architecture SNN accelerators~\cite{guo2023end, khodamoradi2021s2n2}, we have identified the following observations:

\subsubsection{Effectiveness of Input-Weight Pair}

Since matrix–vector multiplication produces zero whenever either operand is zero, zero inputs or weights make the corresponding computations redundant. The main challenge in streaming architectures is therefore to identify valid (non-zero) input–weight pairs with minimal control logic.

\subsubsection{Influence of Data Fetching Order}

During inference, the IFM is dynamic while the kernels (weights) remain fixed. Because all weight values are known in advance, a weight-priority scheme can use precomputed metadata to locate non-zero weights — avoiding real-time decoding, extra decoders, or memory fetches. In contrast, input-priority methods require real-time decoding of IFM metadata to fetch the corresponding weights, which complicates the hardware design.

\subsubsection{Fetching Overhead Difference between Input and Weight}

Although both IFM and weights are stored in local memory (e.g., SRAM or scratchpad memory) in streaming architectures, the IFM is typically in binary format, whereas weights are stored in fixed-point format (e.g., 16-bit) on hardware. Consequently, fetching the IFM is generally more efficient than retrieving the weights.

Building on these observations, we aim to integrate both temporal and spatial sparsity in SNN streaming architectures to enhance efficiency while preserving high throughput.

\section{Sparsity-Aware Output-Channel Dataflow Streaming (SAOCDS) Accelerator}

Given the high throughput requirements of RF tasks, we adopt a streaming architecture accelerator. In light of the opportunities for sparsity utilization identified in the previous section, we propose for the first time a fully-pipelined streaming SNN accelerator that integrates both temporal and spatial sparsity. This design enables automatic load balancing across PEs and delivers high throughput even under resource constraints. In this section, we first analyze the challenges of combining both sparsity types in a streaming architecture. Based on that analysis, we introduce our proposed sparsity-aware accelerator and explain how we embed sparsity-utilization algorithms into its streaming dataflow.

\subsection{Challenges of Combined Temporal–Spatial Sparsity in Streaming Architectures}

When designing streaming architectures, it is common to apply uniform operations across all inputs to the PEs to balance the workload and minimize the need for control logic. This approach intuitively aims to reduce the number of conditional operations—especially nested conditionals—which can lengthen the critical path, slow down basic compute elements, and introduce workload imbalance. However, leveraging sparsity inherently involves skipping unnecessary operations. Thus, a key challenge is how to utilize both temporal and spatial sparsity without introducing nested conditionals.

In input-priority streaming architectures \cite{khodamoradi2021s2n2,guo2023end}, the IFM iteration omits sparsity detection, making real-time exploitation of temporal sparsity unfeasible. Introducing on-the-fly checks to skip zero-valued inputs would both create workload imbalance — due to channel-wise sparsity variation — and require per-stream control logic, which dramatically increases hardware overhead compared to reconfigurable designs like systolic arrays. Consequently, these architectures usually exploit only temporal sparsity during accumulation: input spikes enable or disable the accumulator’s operation, maintaining balanced latency among PEs. Still, IFM fetching and weight selection access all inputs (including zero) and their weights, causing redundant data transfer. As a result, the combined gain from temporal and spatial sparsity remains severely limited, and the overhead of attempting full sparsity integration makes the approach impractical.

An alternative is to adopt weight-priority algorithms. Existing work on GOAP~\cite{lien2022sparse} still relies on a systolic-array architecture enhanced with control logic, uses non-zero weight metadata to locate corresponding inputs. With the assistance of routers and schedulers, GOAP can dispatch only non-zero input–weight pairs to PEs, achieving balanced execution without enforcing uniform data flow.

However, naively transplanting such control logic into a streaming architecture — replicating it for every data path to preserve high parallelism — introduces significant challenges: (1) hardware resource usage and logic latency balloon as control logic is duplicated per path (unlike the shared logic in reconfigurable arrays); (2) parallel paths may contend for memory access, leading to conflicts or incorrect behavior; (3) due to the real-time and unpredictable nature of inputs, varying sparsity patterns across paths cause uneven workload, resulting in latency imbalance and synchronization issues (some PEs overloaded, others idle).

These issues indicate that migrating weight-priority sparsity methods from reconfigurable arrays to streaming architectures is nontrivial. It demands a carefully designed dataflow that — with minimal control logic — preserves high parallelism, balances workload across PEs, and maintains consistent streaming dataflow across layers.

To meet these requirements, we propose a hybrid scheme: precompute and compress non-zero weights (since weights are fixed at inference), then during iteration traverse only non-zero weights; subsequently, apply temporal sparsity during accumulation, using input spikes as enable signals to skip unnecessary operations. This design combines data compression and sparsity-aware skipping while retaining the simplicity and throughput advantages of streaming architectures.

\subsection{Sparsity Utilization in Fully-Connected Layers}

SNNs primarily consist of two types of layers: fully connected (FC) layers and convolutional layers. The FC layer performs matrix-vector multiplication and accumulation operations. To handle these computations, the matrix-vector threshold unit (MVTU) is employed. Each MVTU comprises multiple PEs, with each PE containing several single instruction multiple data (SIMD) lanes to facilitate parallel processing of matrix operations.

In FC layers, each weight corresponds to a unique connection between neurons, and each weight is utilized once per timestep. To optimize computations, it's essential to identify the non-zero input-weight pair. This scenario resembles a logical AND operation, where the computation is necessary only if both operands are active.

\begin{figure}
    \centering
    \includegraphics[width=1\linewidth]{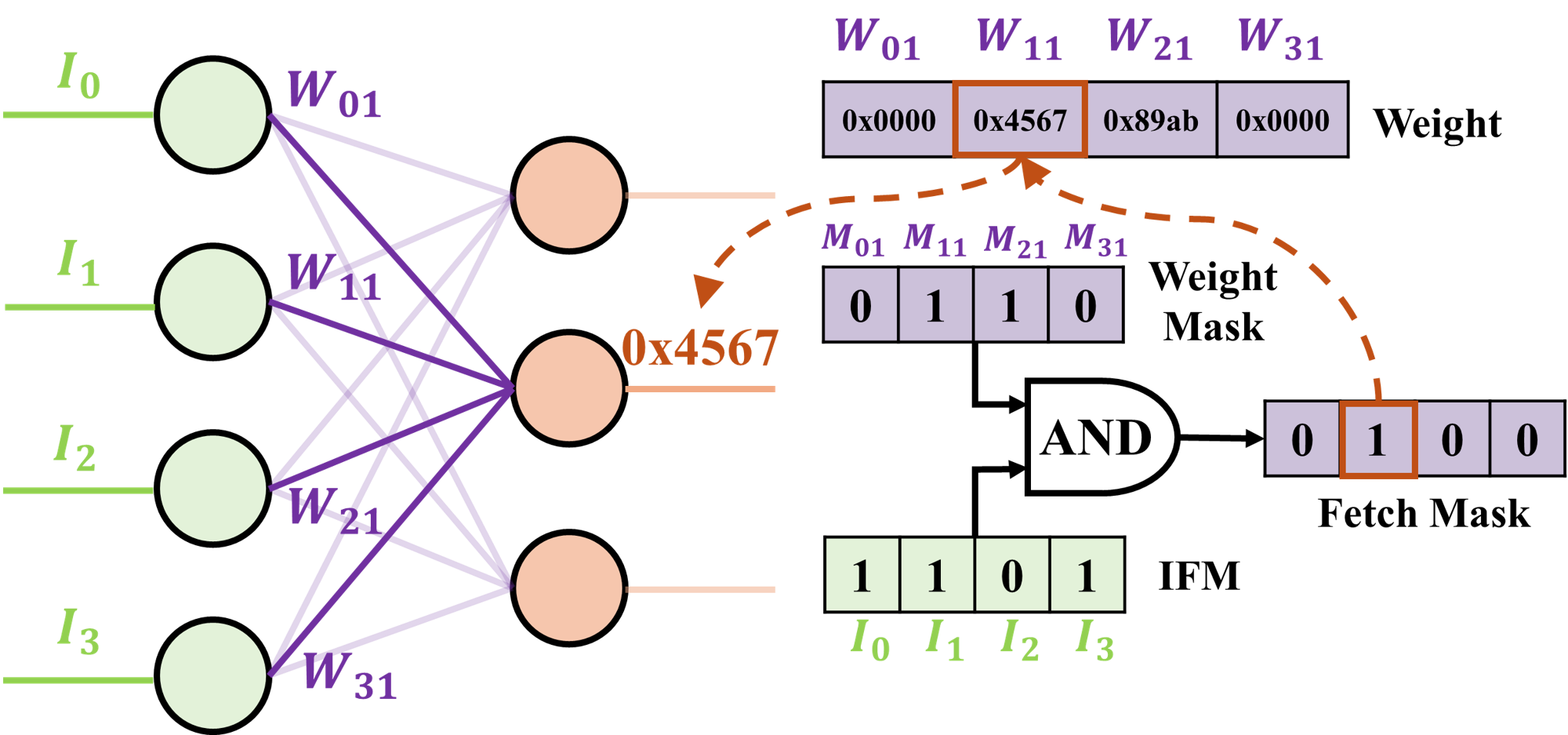}
    \caption{Weight mask for fully-connected (FC) layer.}
    \label{fig: weight mask}
\end{figure}

Weights are typically stored in fixed-point format (e.g., 16-bit). Fetching a weight only to determine if it's zero or not results in unnecessary memory access. Given that kernel weights remain constant during inference, we can preprocess them to identify non-zero values. One effective method is the implementation of a weight mask (WM), as illustrated in Fig. \ref{fig: weight mask}.

In this approach, each weight is associated with a 1-bit mask indicating whether it's non-zero. This results in a storage overhead of only 1/16 compared to the original 16-bit weight storage. When a neuron receives input from the previous layer, the input's binary nature allows it to serve directly as a mask. By performing a logical AND operation between the input and the weight mask, we derive a fetch mask (FM) that identifies which weights need to be fetched and processed.

Compared to traditional methods, where all weights corresponding to active inputs are fetched and processed, this technique significantly reduces computational load. For instance, in Fig.~\ref{fig: weight mask}, the traditional method would fetch and process three weights based on IFM, but including two zeros. In contrast, the WM method fetches and processes only one non-zero weight based on FM, reducing both memory access and computation. The additional hardware requirement is minimal, involving only a 4-bit AND gate, which is straightforward to implement.

This design becomes increasingly beneficial as network complexity grows, offering substantial improvements in efficiency and performance. Similar strategies have been employed in ANNs \cite{hunter2022two} to exploit sparsity in neural networks, leading to reductions in memory usage and computational demands.

\subsection{Sparsity Utilization in Convolutional Layers}

Unlike FC layers, where each weight is used once per timestep, convolutional layers reuse weights across multiple input pixels due to the nature of the convolution operation. While the weight mask technique can still reduce some hardware resource utilization and unnecessary data fetches, it becomes less efficient in this context because of the repeated logical AND operations. Therefore, we need to develop an efficient strategy tailored for convolutional layers to effectively exploit sparsity without incurring significant overhead.

\begin{figure}
    \centering
    \includegraphics[width=0.9\linewidth]{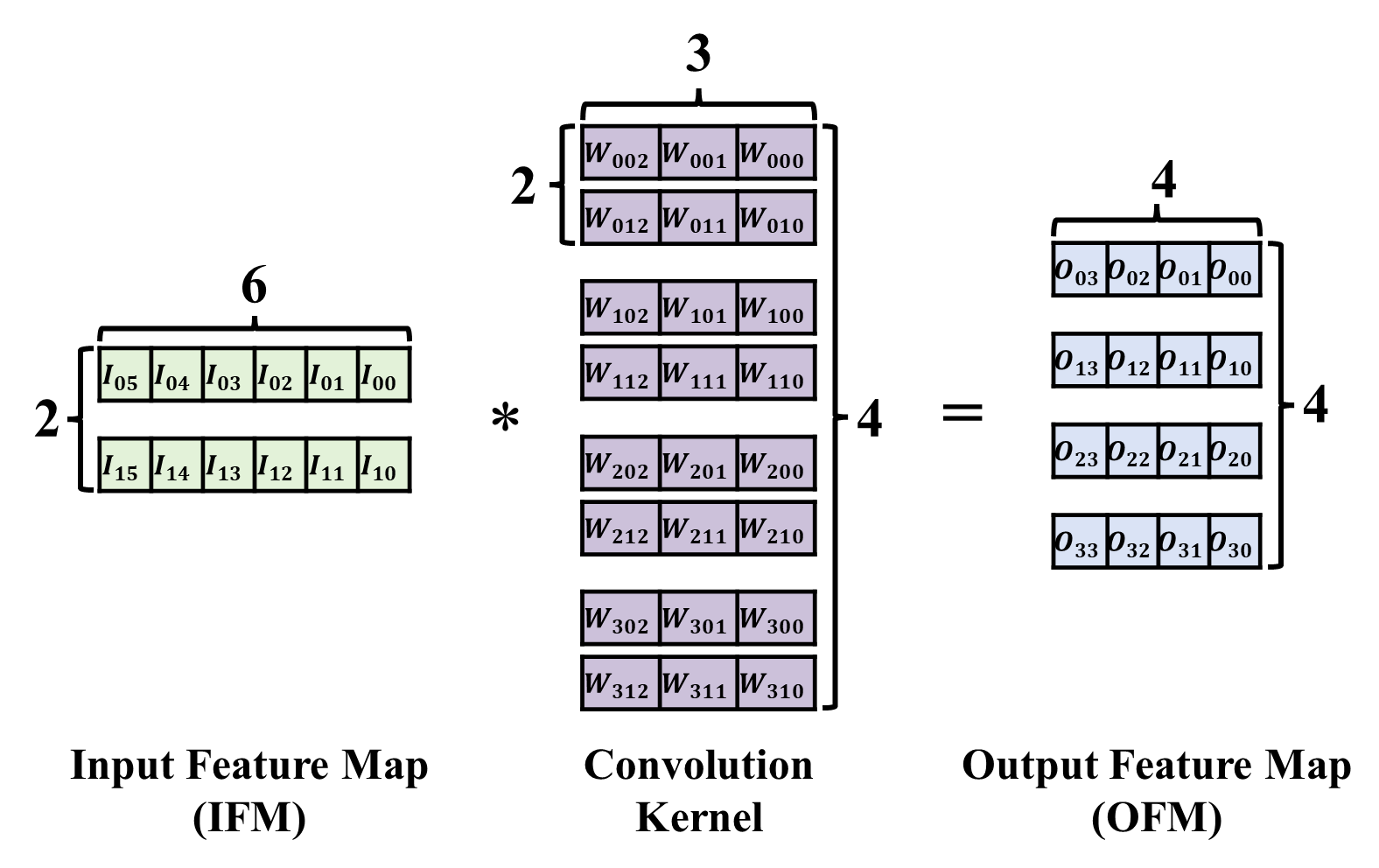}
    \caption{Simple example of a convolution operation}
    \label{fig: convolution example}
\end{figure}

To facilitate the introduction of the subsequent algorithm, we present a simplified example of a convolution operation with the following dimensions, as shown in Fig. \ref{fig: convolution example}.

\begin{itemize}
    \item \textbf{Input Feature Map (IFM):} (1, 6, 2), representing height (H), width (W), and number of input channels (IC), respectively.
    \item \textbf{Kernel:} (1, 3, 2, 4), corresponding to height (H), width (W), number of input channels (IC), and number of output channels (OC).
    \item \textbf{Output Feature Map (OFM):} (1, 4, 4), denoting height (H), width (W), and number of output channels (OC).
\end{itemize}

In this example, the IFM dimensions are specified after padding, and the stride is set to 1. The heights of the IFM, OFM, and kernels are all set to 1, aligning with the characteristics of long input arrays commonly found in RF input signals.

\subsubsection{Gated One-to-All Product Algorithm}

Building upon our previous observations, we recognize that during inference, kernel weights remain fixed, and fetching based on weight priority is more efficient, as the data in IFM typically has a smaller size. Therefore, utilizing the fixed sparse weights to identify corresponding IFM entries and determining whether the input is zero to decide on accumulation is a viable solution.

\begin{figure*}[t]
  \centering
  \includegraphics[width=\textwidth]{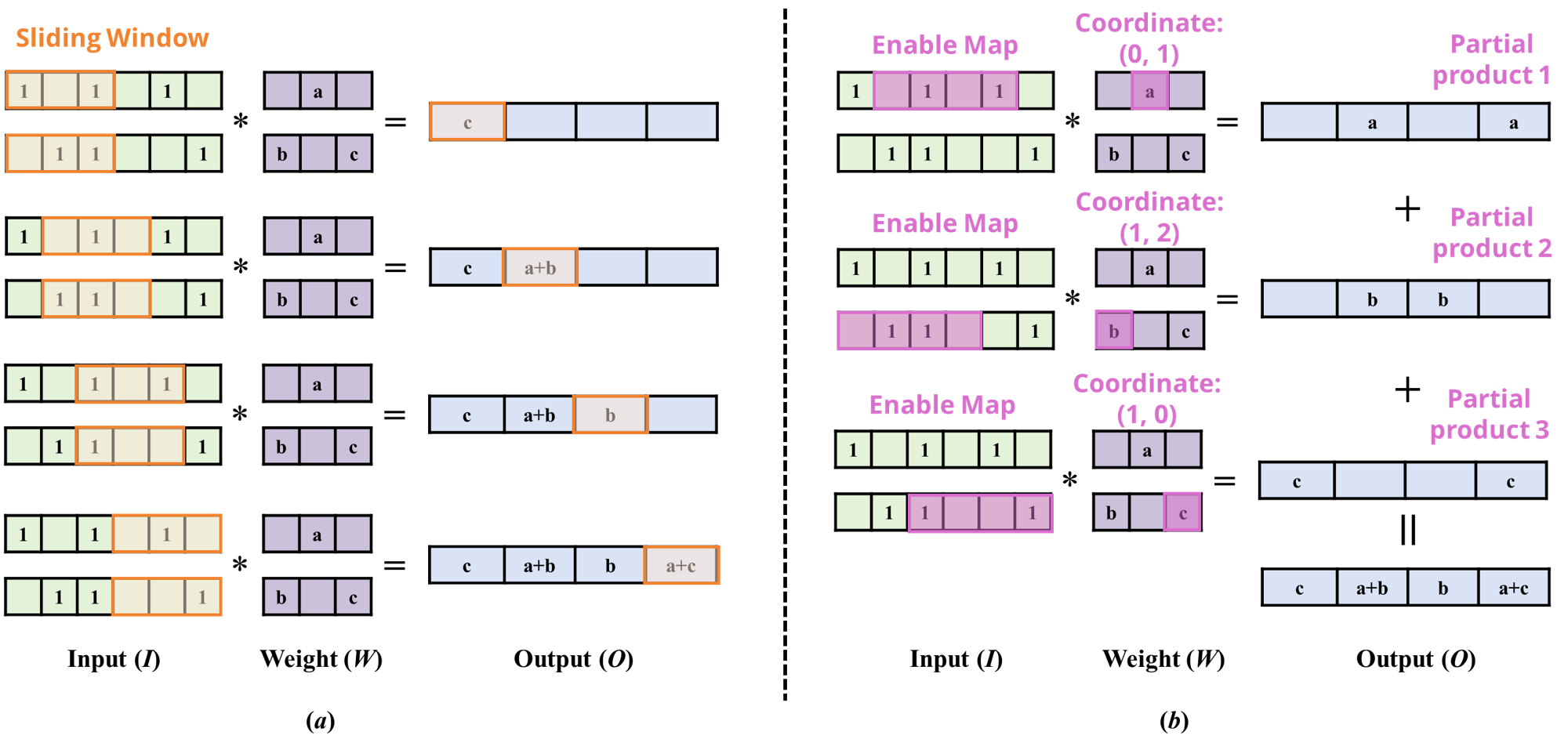}
  \caption{Comparison of working principle of (a) sliding window (SW) and (b) gated one-to-all product (GOAP) algorithm of a single channel in a convolutional layer}
  \label{fig: comparison of SW and GOAP}
\end{figure*}

This method is referred to as the gated one-to-all product (GOAP) algorithm~\cite{lien2022sparse}

\begin{table}[htbp]
\centering
\caption{Comparison between Sliding Window (SW) method and Gated One-to-All Product (GOAP) method in terms of the number of input fetches, weight fetches, and accumulations.}
\label{tab:fetch_compare_2}
\renewcommand{\arraystretch}{1.6}
\setlength{\tabcolsep}{4pt}
\begin{tabular}{|c|c|c|}
\hline
\textbf{Method} & \textbf{SW} & \textbf{GOAP} \\
\hline
\# Input Fetch & $3 \times 2 \times 4 = 24$ & $4 \times 1 \times 3 \times 4 = 48$ \\
\hline
\# Weight Fetch & $3 \times 2 \times 4 \times 4 = 96$ & $3 \times 4 = 12$ \\
\hline
\# Accumulation & $(4 + 3 + 3 + 2) \times 4 = 48$ & $(2 + 2 + 2)\times4 = 24$ \\
\hline
\end{tabular}
\end{table}

Unlike the sliding window (SW) method, the GOAP method focuses solely on the non-zero weights in the kernel. It iterates through these non-zero weights and then applies IFM fetching. Consequently, the generation of the output dataflow is also modified.

\subsubsection{Dataflow Comparison between SW and GOAP Methods}

Based on the configuration of the accelerator, each PE automatically retrieves the membrane potential from the previous timestep from memory and applies the decay factor ($\alpha$) to model the leaky behavior characteristic of LIF neurons. Subsequently, the IFM stream is fed into each PE, which then fetches the corresponding synaptic weights from memory. The PE performs the weighted accumulation of inputs, and upon reaching a predefined threshold, determines whether the neuron should fire. If so, the potential will be reset softly. After this process, the updated membrane potential for the current timestep is written back to memory. The MVTU module encapsulates the functionalities of matrix-vector multiplication, accumulation, and the emulation of the LIF neuron model, thereby streamlining the computational pipeline.

Consider the convolution example in Fig.~\ref{fig: convolution example}, the working principles of the SW and GOAP methods for a single channel are illustrated in Fig.~\ref{fig: comparison of SW and GOAP}, highlighting their differences. For a fair comparison, both temporal and spatial sparsity are set to 50\%, and all four kernels are assumed to have identical weight distributions for simplicity.

In Fig.~\ref {fig: comparison of SW and GOAP}(a), the PE receives a stream of IFMs with dimensions $(H, W, IC) = (1, 3, 2)$. It fetches the corresponding weights of the same size based on the index and accumulates over the input channels to compute a single output pixel. This operation is performed in parallel across all output channels. The output pixel then slides pixel by pixel with the movement of the sliding window. Since each kernel is fetched four times with sliding window moves, a total of $3 \times 2 \times 4 = 24$ weight fetches are performed for one output channel, and 96 weight fetches for all channels. Similarly, with four sliding windows of the same size, the inputs are fetched 24 times. As the IFM is used by all the output channels in parallel, the total fetch time is still 24. Regarding accumulation, as the SW algorithm only exploits temporal sparsity—accumulating whenever the IFM is non-zero—there are $4 + 3 + 3 + 2 = 12$ accumulations for one output channel. The total number of input fetches, weight fetches, and accumulations for the convolution operation shown in Fig.~\ref {fig: convolution example} are summarized in Table~\ref{tab:fetch_compare_2}.

Conversely, Fig.~\ref{fig: comparison of SW and GOAP}(b) illustrates the GOAP method. For instance, consider a non-zero weight $a$ located at coordinate $(0, 1)$, indicating the first input channel and the second horizontal position. Analyzing the sliding window operation reveals that only IFM values at positions 1–4 in the first input channel will be multiplied with the weight $a$. This range is referred to as the weight $a$'s enable map (EM). Thus, each non-zero weight is fetched once, and its EM fetches four corresponding input values. Considering temporal sparsity, only two accumulations are executed for weight $a$. Similarly, weights $b$ and $c$ also result in two accumulations each. Consequently, for one output channel, the GOAP algorithm performs 12 input fetches, 3 weight fetches, and 6 accumulations. The total number of input fetches, weight fetches, and accumulations across all output channels is summarized in Table~\ref{tab:fetch_compare_2}.

Comparing these results, the GOAP method fetches input data twice as often as the SW method. However, it requires only one-eighth the number of weight fetches and performs half the number of accumulations. This efficiency arises from GOAP's utilization of both temporal and spatial sparsity, in contrast to SW's reliance solely on temporal sparsity. Notably, in this design, each input has a bit-length of 1, whereas each weight has a bit-length of 16. Consequently, the total data fetched by SW amounts to 1,560 bits, while GOAP fetches only 240 bits—merely 15.4\% of SW's total. Coupled with the reduced number of accumulations, the GOAP method significantly conserves operations and power related to data fetching and accumulation.

\begin{algorithm}
\caption{Algorithm of the Sparsity-Aware Output Channel Dataflow Streaming (SAOCDS) SNN Accelerator for One Convolutional Layer.}
\label{alg: SAOCDS}
\begin{algorithmic}[1]

\Statex \textbf{Input:} IFM streamed sequentially by the input channel.
\Statex \textbf{Output:} OFM generated sequentially by the output channel, forwarded to the next layer.
\For{$t = 0$ to $T - 1$}
    \State $IC\_read \gets 0$
    \State $pre\_oc \gets OC$
    \State $oc \gets 0$
    \State $next\_oc \gets 0$
    \For{$nnz = 0$ to $NNZ - 1$}
        \State $oc \gets \left\lfloor \frac{W[nnz].RI}{IC} \right\rfloor$
        \State $next\_oc \gets \left\lfloor \frac{W[nnz + 1].RI}{IC} \right\rfloor$
        \State $ic \gets W[nnz].RI \% IC$
        \If{$IC\_read < IC$}
            \State Input $I[ic]$
            \State $IC\_read \gets IC\_read + 1$
        \EndIf
        \If{$oc \neq pre\_oc$}
            \State Load $V[oc]$
            \State Decay $V[oc]$
        \EndIf
        \For{$oi = 0$ to $OI - 1$}
            \If{$I[ic][oi + W[nnz].CI] == 1$}
                \State $V[oc][oi] \gets V[oc][oi] + W[nnz].D$
            \EndIf
        \EndFor
        \If{$next\_oc \neq oc$}
            \State Output $O[oc]$
            \State Store $V[oc]$
        \EndIf
        \State $pre\_oc \gets oc$
    \EndFor
\EndFor
\end{algorithmic}
\end{algorithm}

\subsubsection{Compressed Format of Weights}

Incorporating weight sparsity into neural network design necessitates selecting an appropriate storage format for sparse data. Common formats include coordinate (COO), and compressed sparse row (CSR). Each of these formats utilizes three arrays to store non-zero values and their corresponding metadata.

In the COO format, the first array stores the non-zero values, the second records the row indices of these values, and the third contains the column indices. Conversely, the CSR format's first array holds the non-zero values, the second array (known as the index pointer) indicates the starting position of each row in the data and index arrays, and the third stores the column indices of the non-zero values.

\begin{figure}
    \centering
    \includegraphics[width=1\linewidth]{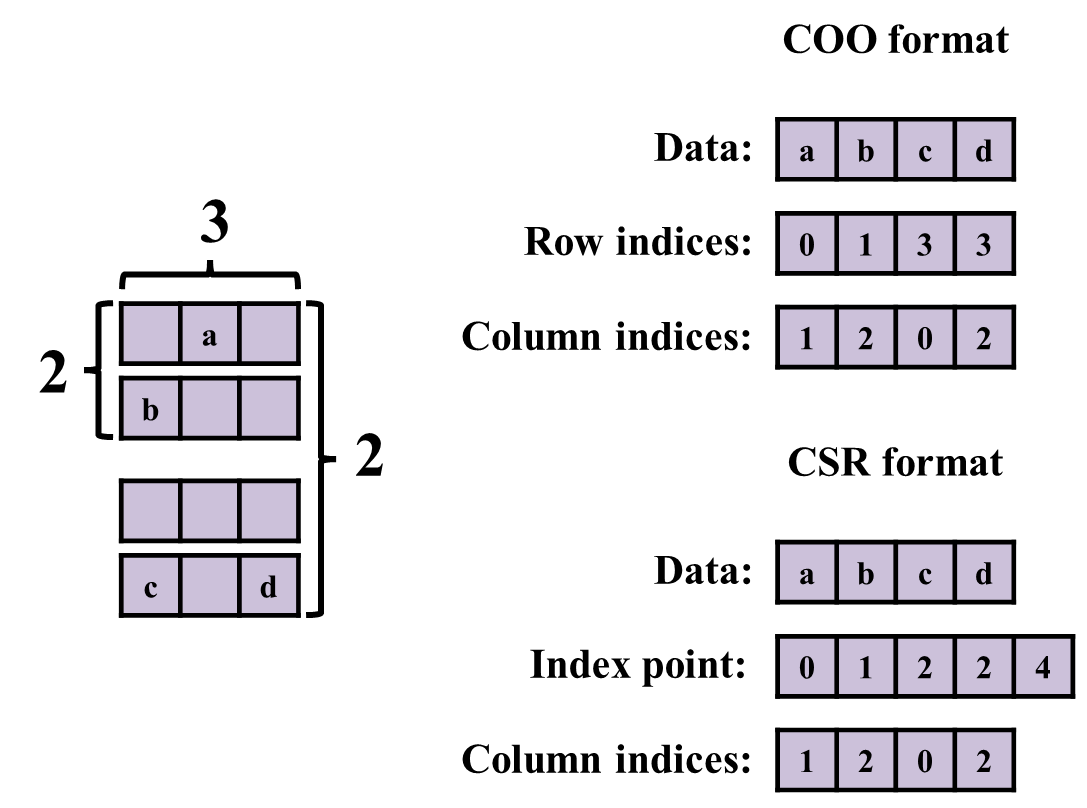}
    \caption{Example of sparse matrix storage in coordinate (COO) and compressed sparse row (CSR) format.}
    \label{fig: example of compressed format}
\end{figure}

Consider the example in Fig. \ref{fig: example of compressed format}, which illustrates a kernel with dimensions $(H, W, IC, OC) = (1, 3, 2, 2)$ stored in both COO and CSR formats. While these compressed formats are designed for 2D matrices, our kernel possesses four dimensions. Although $H = 1$ in this example, resulting in three dimensions, storing the compressed matrix of each output channel as a separate component poses challenges in hardware implementation. Specifically, fine-grained memory separation is difficult, especially when the number of non-zero values varies across output channels. Hardware typically allocates space based on the largest component, potentially leading to space inefficiencies. To address this, we merge the input and output channel indices into the row indices, effectively transforming the 3D tensor into a 2D matrix.

\begin{equation}
    \label{eq: IC index calcu}
    IC\_index = W.RI \;\%\;\#IC
\end{equation}

\begin{equation}
    \label{eq: OC index calcu}
    OC\_index = W.RI \;/\;\#IC
\end{equation}

The COO format offers enhanced flexibility by allowing direct access to row and column indices without iterating through adjacent rows, simplifying control logic. Therefore, we adopt the COO format for our design. Each non-zero weight data($W.D$) is associated with a row index($W.RI$) and a column index($W.CI$). The input channel index (IC\_index) and output channel index (OC\_index) can be calculated with (\ref{eq: IC index calcu})(\ref{eq: OC index calcu}). The column index directly corresponds to the column index of the non-zero weight in the input channel.

To assess the overhead of COO storage, Table \ref{tab: COO overhead} lists the bit-widths of $W.D$, $W.RI$, and $W.CI$, as well as the “Total Length” per non-zero entry. The “Amount” column represents the total number of weights per layer under dense (100 \%) storage.

For example, in Layer 2 the convolution filter size is $1 \times 11 \times16 \times32$, yielding 5,632 weights in dense storage (16 bits each). In the sparse COO format, each non-zero weight is encoded with a 9-bit row index and a 4-bit column index, totaling 29 bits. Thus, at sparsity density $X$, the required storage is $29 \times 5632 \times X = 163328X ~bits$. Comparing this to dense storage indicates a “break-even” density of approximately 55.2\% for Layer 2; when actual density is below this threshold, sparse storage becomes more bit-efficient. Similar calculations for other layers are summarized in the table.

In practice, however, on-chip memory allocation constraints may erode these savings. For our target FPGA, BRAM words must align to supported widths (e.g., 1, 2, 4, 9, 18, 36 bits), and the smallest allocatable unit is half a BRAM tile ($18 kbits$). Consequently, sparse weight data are often padded or rounded up to the nearest allocation block. Since we did not apply custom bit-packing or index-concatenation (to avoid added decoding logic and latency), the theoretical memory savings from sparsity may be diminished or even eliminated under moderate density.

\begin{table}[htbp]
\centering
\caption{COO overhead vs. Dense storage comparison of three convolutional layers.}
\label{tab: COO overhead}
\renewcommand{\arraystretch}{1.6}
\setlength{\tabcolsep}{1.8pt}
\begin{tabular}{|c|c|c|c|c|c|c|c|c|c|}
\hline
\makecell[c]{\textbf{Layer} \\ \textbf{No.}} & 
\makecell[c]{\textbf{W.} \\ \textbf{D}} & 
\makecell[c]{\textbf{W.} \\ \textbf{RI}} & 
\makecell[c]{\textbf{W.} \\ \textbf{CI}} & 
\makecell[c]{\textbf{Total} \\ \textbf{Length}} & 
\textbf{Amount} & 
\textbf{Density} & 
\makecell[c]{\textbf{Dense} \\ \textbf{Total} \\ \textbf{Bit}} & 
\makecell[c]{\textbf{COO} \\ \textbf{Total} \\ \textbf{Bit}} & 
\makecell[c]{\textbf{Break} \\ \textbf{-Even} \\ \textbf{Density}} \\
\hline
L1 & 16 & 5 & 4 & 25 & 352 & $X$ & 5632 & $8800X$ & 64.00\% \\
\hline
L2 & 16 & 9 & 4 & 29 & 5632 & $X$ & 90112 & $163328X$ & 55.17\% \\
\hline
L3 & 16 & 11 & 3 & 30 & 10240 & $X$ & 163840 & $307200X$ & 53.33\% \\
\hline
\end{tabular}
\end{table}

\subsubsection{Sparsity-Aware Output Channel Dataflow Streaming (SAOCDS) Algorithm}

\begin{figure*}[t]
  \centering
  \includegraphics[width=\textwidth]{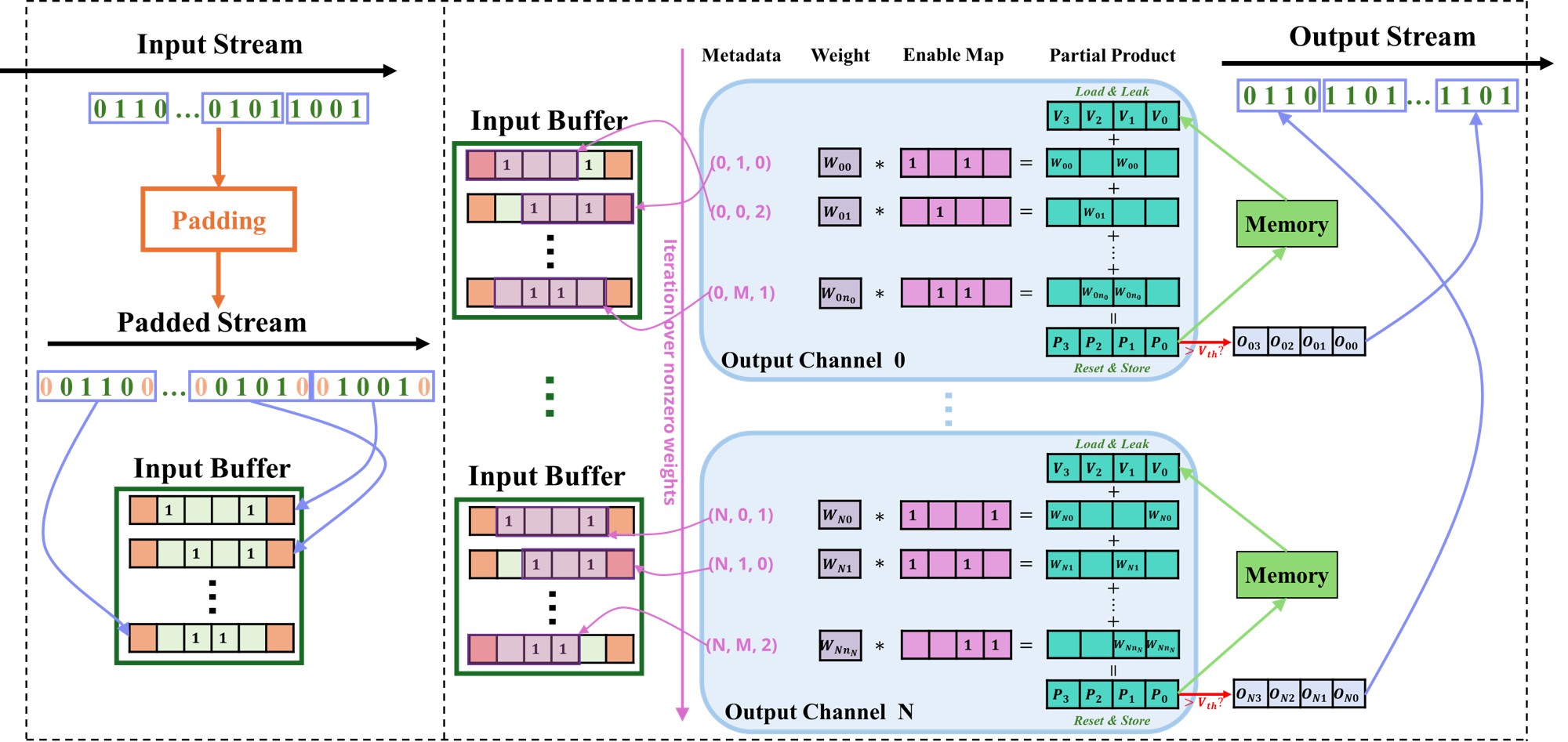}
  \caption{Dataflow of a single layer in the SAOCDS accelerator}
  \label{fig: working principle of SAOCDS}
\end{figure*}

Based on the dataflow analysis of the GOAP method and compressed format, we can derive out sparsity-aware output channel dataflow streaming (SAOCDS) algorithm. The dataflow is shown in Fig. \ref{fig: working principle of SAOCDS}.

Within each accelerator layer, we proceed as follows. First, we read spike signals from the previous layer in input-channel order, then pad and store them in an input buffer to reuse in subsequent iterations (as shown on the left side of Fig. \ref{fig: working principle of SAOCDS}). Then, we iterate over all non-zero weights in output-channel order, using the weight data stored in COO format and its metadata to perform accumulation. Finally, we compute the membrane potentials and generate output spikes (as depicted on the right side of Fig. \ref{fig: working principle of SAOCDS}). The detailed iteration steps and variable computations are given in the Algorithm.\ref{alg: SAOCDS}.

For each time step, we begin by initializing the iteration variables (lines 2–5). We then iterate over the non-zero weights stored in compressed format (line 6). For each weight, we decode the associated output channel index $oc$ (line 7), the index of the next output channel $next\_oc$ (line 8), and the input channel index $ic$ (line 9). If the current input channel index is less than the total number of input channels, we read a new input channel's data from the input buffer (lines 10–13). If the current $oc$ differs from the previous one, this indicates a new output channel, and we must read the membrane potential (state) of the new corresponding neuron (lines 14–17). Next, we iterate over the output pixels $oi$ within the enable map of this weight. If the corresponding value is $1$, we accumulate the partial product into the neuron’s membrane potential; otherwise, we skip the accumulation (lines 18–22). Note that the enable map has the same dimensions as a single output channel, and each non-zero weight contributes a single partial product. After processing a non-zero weight, if $next\_oc$ is not equal to $oc$, it indicates the end of the accumulation phase for the current output channel. At this point, the neuron’s potential is compared with its threshold to determine whether it emits a spike. If so, a spike is output to this pixel, and a soft reset is performed. Then the updated membrane potential is written back to memory (lines 23–26). Finally, the previous output channel index $pre\_OC$ is updated (line 27), completing one iteration of a non-zero weight (lines 6–28). Once all non-zero weights for the current time step have been processed, the computation for that step concludes, and we advance to the next timestep.

By iterating over non-zero weights, we exploit spatial sparsity without requiring additional control logic or real-time decoders to retrieve metadata. Utilizing input signals to gate accumulation allows us to leverage temporal sparsity by skipping unnecessary computations. Furthermore, adopting an output-channel dataflow—where each layer transfers a single output channel at a time in a fixed order following the output channel indices—enables direct connection between the outputs of the previous layer and the inputs of the next layer without any extra control logic. Additionally, since the accumulation dimension for each non-zero weight matches the output channel's dimension (also the dimension of the EM), the workload is inherently balanced. Moreover, as the neuron state is only loaded and stored during output channel transitions, frequent memory accesses within the same output channel are eliminated. In summary, our streaming dataflow integrates both temporal and spatial sparsity without extra control logic, ensuring balanced workload distribution for each non-zero weight.

\subsection{Supplement Algorithm of the Sparsity}

Incorporating sparsity into our design enhances computational efficiency but introduces potential logical challenges during iteration execution. High sparsity levels can lead to specific issues that require careful handling.

\begin{algorithm}
\caption{SAOCDS Algorithm with Supplementary Sparsity-Handling Mechanisms for One Convolutional Layer.}
\label{alg: SAOCDS with extra iteration}
\begin{algorithmic}[1]
\Statex \textbf{Input:} IFM streamed sequentially by input channel.
\Statex \textbf{Output:} OFM generated sequentially by output channel, forwarded to the next layer.
\State $REPS \gets NNZ + \text{\#extra\_I} + \text{\#empty\_I}$
\For{$t = 0$ to $T - 1$}
    \State $IC\_read \gets 0$
    \State $pre\_oc \gets OC$
    \State $oc \gets 0$
    \State $nnz \gets 0$
    \For{$reps = 0$ to $REPS - 1$}
        \State $nnz\_oc \gets \left\lfloor \frac{W[nnz].RI}{IC} \right\rfloor$
        \State $nnz\_next\_oc \gets \left\lfloor \frac{W[nnz + 1].RI}{IC} \right\rfloor$
        \If{$IC\_read < IC$}
            \State Input $I[ic]$
            \State $IC\_read \gets IC\_read + 1$
        \EndIf
        \If{$oc \neq nnz\_oc$}
            \State Load $V[oc]$
            \State Decay $V[oc]$
            \State Output $O[oc]$
            \State Store $V[oc]$
            \State $oc \gets oc + 1$
        \Else
            \State $ic \gets W[nnz].RI \% IC$
            \If{$ic < IC\_read$}
                \If{$oc \neq pre\_oc$}
                    \State Load $V[oc]$
                    \State Decay $V[oc]$
                \EndIf
                \For{$oi = 0$ to $OI - 1$}
                    \If{$I[ic][oi + W[nnz].CI] == 1$}
                        \State $V[oc][oi] \gets V[oc][oi] + W[nnz].D$
                    \EndIf
                \EndFor
                \If{$nnz\_next\_oc \neq oc$}
                    \State Output $O[oc]$
                    \State Store $V[oc]$
                    \State $pre\_oc \gets oc$
                    \State $oc \gets oc + 1$
                \Else
                    \State $pre\_oc \gets oc$
                \EndIf
                \State $nnz \gets nnz + 1$
            \EndIf
        \EndIf
    \EndFor
\EndFor
\end{algorithmic}
\end{algorithm}

\subsubsection{Empty Iteration}

In our architecture, the previous layer outputs data channel by channel. However, the kernel may lack non-zero weights for certain input channels. For instance, consider a kernel that has non-zero weights only for input channels 0 and 2. During the first iteration, the system reads data from input channel 0, and since corresponding non-zero weights exist, computations proceed as expected. In the second iteration, the system reads data from input channel 1, but the kernel lacks non-zero weights for this channel. Moreover, the non-zero weights correspond to input channel 2, whose data has not yet been updated or is unavailable (In Algorithm.\ref{alg: SAOCDS}, $IC\_read = 1$ while $ic = 2$). Attempting computations at this point could lead to logical errors. To mitigate this, we designate such iterations as "empty iterations", during which no computations are performed. The system resumes operations once the necessary input data becomes available (Algorithm.\ref{alg: SAOCDS with extra iteration}, conditional statement in line 22).

Note that empty iterations occur only during the computation of the first output channel. For subsequent output channels, input data from all channels is reused, preventing this issue.

\subsubsection{Extra Iteration}

Our algorithm iterates over non-zero values in the kernels. However, high sparsity may result in certain output channels lacking non-zero weights, especially when the dimension of the output channel is small. In Algorithm.\ref{alg: SAOCDS}, when transitioning from one output channel to another, the current output channel's state is written back to the state variable arrays, and the next output channel's state is loaded and decayed before accumulation. If an output channel has no nonzero weights, it will not be processed, and its state will not be decayed, leading to inconsistencies.

To address this, we introduce "extra iterations" to update the state and output values of such channels. It is important to decouple the relationship between the iteration over stored non-zero weights and the iteration within the streaming accelerator. The modified algorithm operates as follows (Algorithm.~\ref{alg: SAOCDS with extra iteration}, line 14-19):

\begin{itemize} \item Let the current output channel be $oc$, and the output channel of the current non-zero weights be $nnz\_oc$.
\item If $nnz\_oc = oc$, proceed with the standard iteration.

\item If $nnz\_oc \neq oc$, perform extra iterations for output channel $oc$:
\begin{itemize}
    \item Load the state of the channel.

    \item Apply decay to the state.

    \item Write back the updated state.

    \item Output the value of the intermediate channel.

    \item $oc = oc + 1$
\end{itemize}
\end{itemize}

This approach ensures that all output channels, regardless of sparsity, have their states appropriately updated and outputs generated, maintaining consistency across the network.

It's crucial to highlight that, irrespective of the number of empty or extra iterations required, the previous layer consistently outputs the feature map channel by channel. Given the fixed distribution of our kernel, we can predict the occurrence and location of empty or extra iterations, allowing us to calculate the total number of iterations in advance (Algorithm.\ref{alg: SAOCDS with extra iteration}).

The complete algorithm, incorporating the supplementary sparsity-handling mechanisms, is presented in Algorithm.\ref{alg: SAOCDS with extra iteration}. The empty and extra iterations serve to guarantee correct dataflow under extreme sparsity, while their overhead remains minimal. When sparsity is below 90\%, these iterations number fewer than ten; at 95\% sparsity, they constitute under 10\% of all iterations, and the total iteration count drops to roughly 5\% of that under full density. In summary, we resolve potential sparsity-induced issues with nearly negligible overhead.

\section{AMC system and SNN model}
In this section, we present the trained SNN model used to construct the AMC system, report its classification results, and describe the subsequent model compression and sparsity enhancement processes.

\subsection{Dataset}

The functionality of the AMC system is verified using the RadioML 2016.10A dataset~\cite{o2016radio}, a widely adopted benchmark in AMC research~\cite{o2016radio, jung2022chip, wang2024mr}. This synthetic dataset, generated via GNU Radio, encompasses 11 modulation schemes—comprising 8 digital and 3 analog types—across a signal-to-noise ratio (SNR) range from -20 dB to 18 dB in 2 dB increments. Each sample consists of a 2×128 matrix representing the in-phase (I) and quadrature (Q) components of the RF signal.

To adapt these samples for SNN processing, we employ the Sigma-Delta-based conversion scheme as described in~\cite{guo2023end}. This approach oversamples the original floating-point IQ data and applies a low-pass filter, resulting in a binary output with dimensions (2, 128×OSR), where OSR denotes the oversampling ratio. The binary data is then reshaped into a (2, 128, OSR) tensor, allowing the SNN to process one frame per timestep with an input size of (2, 128) over T = OSR timesteps.

\subsection{SNN Model and Software Classification Results}

\begin{equation}
\label{LIF model}
    \left\{\begin{matrix}
U_t = \alpha \cdot U_{t-1} + W \cdot I - \theta \cdot S_{t-1}\\ 
\left\{\begin{matrix}
S_t = 1,\ \   if \ \   U_t > U_{th0}\\ 
S_t = 0,\ \   if \ \   U_t \leq U_{th0}
\end{matrix}\right.

\end{matrix}\right.
\end{equation}

\begin{figure}
    \centering
    \includegraphics[width=0.9\linewidth]{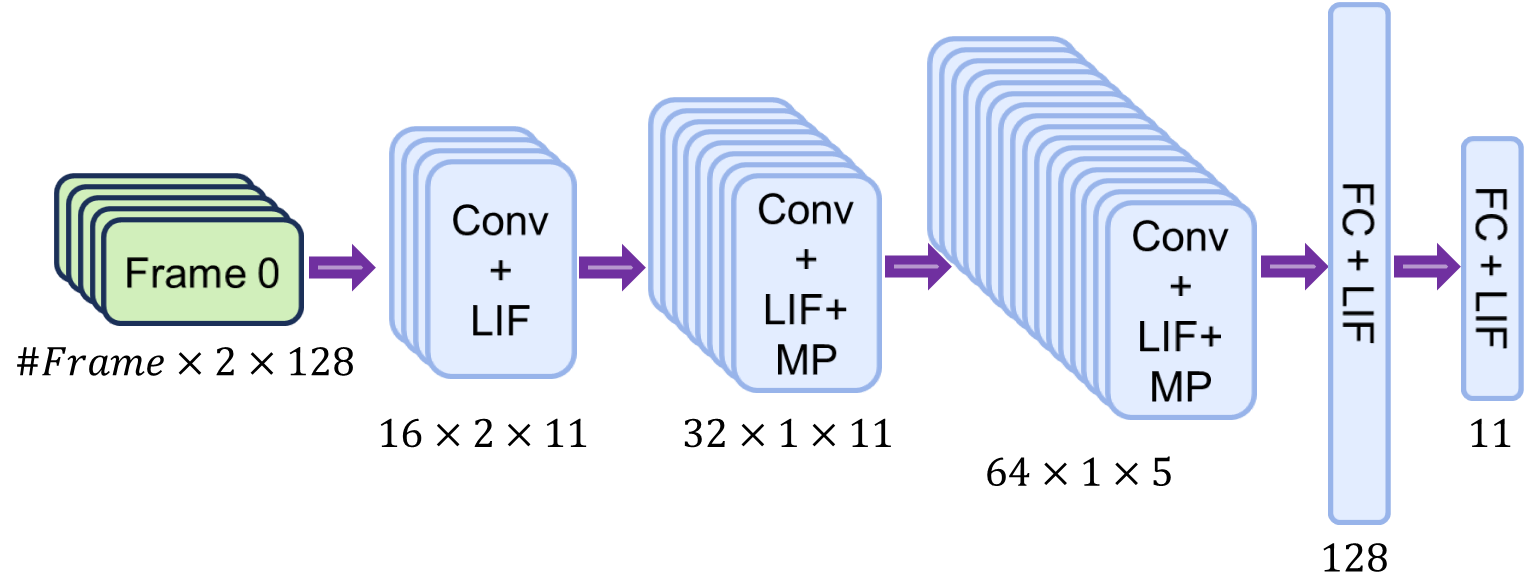}
    \caption{Architecture of the SNN classifier. Abbreviations used in the figure: $Conv$ denotes a convolutional layer, $MP$ denotes a max pooling layer, $FC$ denotes a fully connected layer, and $LIF$ denotes a leaky integrate-and-fire neuron model.}
    \label{fig: SNN architecture}
\end{figure}

The architecture of the SNN classifier model is illustrated in Fig. \ref{fig: SNN architecture} and implemented using the PyTorch framework. Detailed simulation settings are provided in \cite{guo2023end}. The neuron dynamics are modeled using the leaky integrate-and-fire (LIF) model, as defined in Equation~\ref{LIF model}, where $U_t$ denotes the membrane potential at timestep $t$, $\alpha$ is the decay factor, $\theta$ is the soft-reset factor, $W$ is the weight matrix, $S$ represents the spike function, and $U_{th0}$ is the neuron's firing threshold. At each timestep, the membrane potential decays proportionally to the factor $\alpha$. If a spike was generated at the previous timestep, the potential is soft-reset by subtracting $\theta$. The membrane potential then accumulates the product of IFM and kernel, determining whether a spike is emitted in the current timestep. To ensure accurate hardware implementation on an FPGA, $\alpha$, $\theta$, and $U_{th0}$ are treated as trainable parameters for each neuron.

\begin{figure}
    \centering
    \includegraphics[width=1\linewidth]{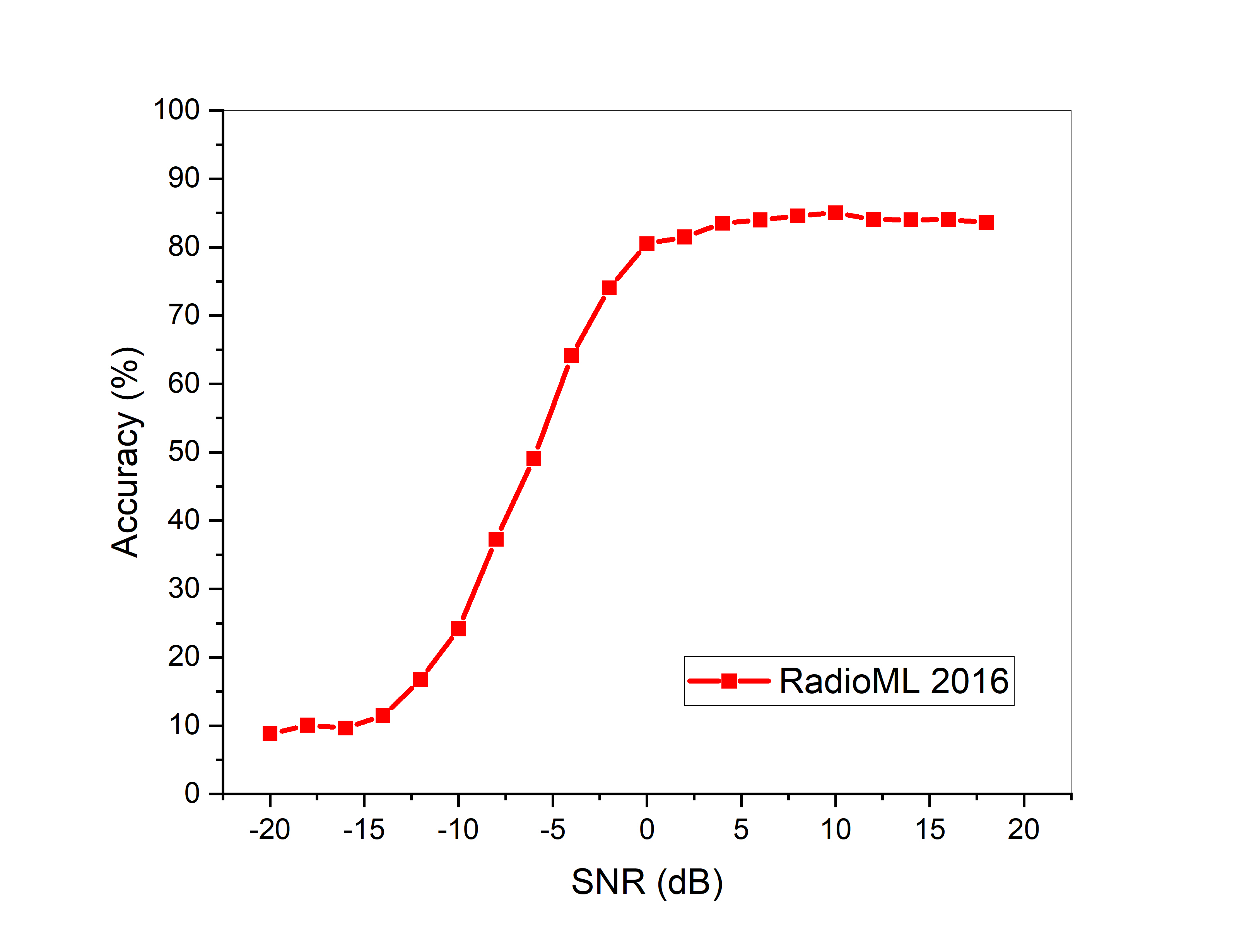}
    \caption{Classification accuracy vs. different Signal-to-Noise Ratio (SNR) on RadioML 2016}
    \label{fig: classification accuracy}
\end{figure}

Figure~\ref{fig: classification accuracy} illustrates the classification accuracy of our model across varying SNRs. At SNRs below 0 dB, the accuracy is relatively low, resulting in an average accuracy of approximately 57\%. However, for SNRs above 0 dB, the model consistently achieves accuracies exceeding 80\%, peaking at around 85\%. These results are comparable to those reported in previous studies on the same dataset~\cite{wang2024mr, o2016radio, pijackova2021radio}.

\subsection{Pruning and Quantization}

\begin{figure*}[t]
  \centering
  \includegraphics[width=\textwidth]{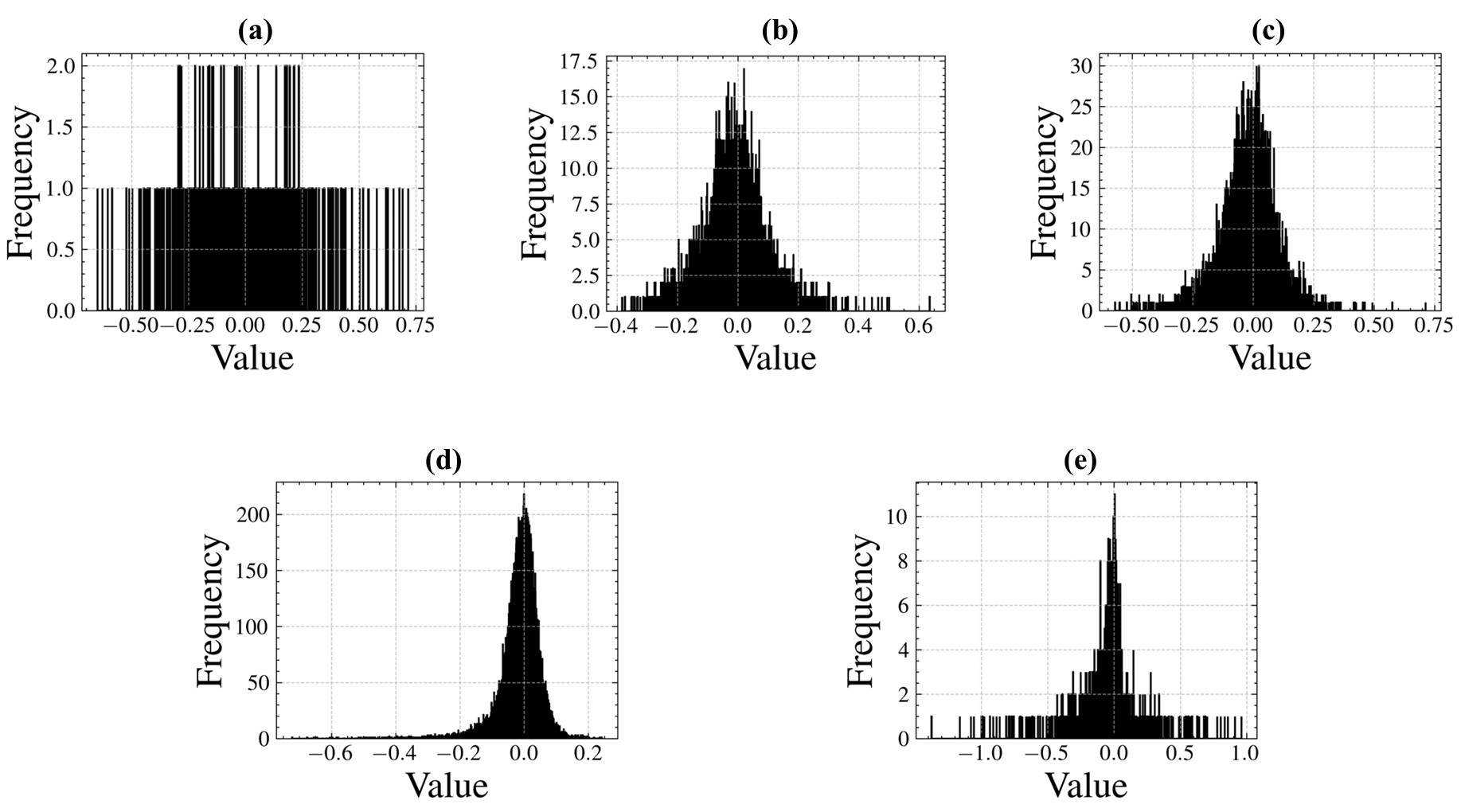}
  \caption{Weight value distributions across layers. (a)-(e) distribution of weight values for layers 1 through 5, respectively. In each histogram, "Frequency" denotes the number of weights falling within each bin of the value range.}
  \label{fig: weight distribution}
\end{figure*}

Fig. \ref{fig: weight distribution} illustrates the absolute value distribution of weights across all layers of our original trained model. To optimize the model for hardware implementation, we apply pruning and quantization techniques to reduce its size and computational complexity.

\subsubsection{Pruning}

Pruning is an effective technique for model compression. We employ fine-grained pruning \cite{han2015learning}, which involves setting individual weights to zero without altering the network's architecture. This approach maintains the model's structure while reducing the number of active parameters.

Analyzing Fig. \ref{fig: weight distribution}, we observe that, except for the first layer, all layers exhibit a significant concentration of weights near zero. This distribution indicates a high potential for pruning without substantial accuracy loss. The first layer shows a less pronounced peak near zero, likely due to its limited number of weights.

To compress the model while maintaining classification accuracy, we employ \textit{L1 unstructured pruning}, which removes weights with the smallest absolute values. Specifically, we adopt a three-phase training schedule over 100 epochs: the initial 20\% epochs are dedicated to learning fundamental features; the subsequent 60\% epochs involve iterative pruning of less significant weights to reduce model size; and the final 20\% epochs focus on fine-tuning to restore any potential performance degradation caused by pruning. This approach aligns with established practices in model compression, where pruning is integrated during training to balance efficiency and accuracy.

\subsubsection{Quantization}

In PyTorch, model weights are stored in 32-bit floating point, providing more precision than required for inference. To optimize FPGA deployment, we quantize these weights to 16-bit fixed point. To prevent accuracy degradation, we employ Learned Step Size Quantization (LSQ) during training—a quantization-aware method that treats the quantization step size as a trainable parameter, optimized via backpropagation using straight-through estimators.

Training with LSQ involves simulating 16-bit quantization during both forward and backward passes, while maintaining full-precision weights for gradient updates. After convergence, we apply the learned step sizes: each weight is scaled, clipped to the 16-bit range, and rounded to its fixed-point representation. The final quantized weights are then stored on the FPGA and used in inference with integer arithmetic. This approach preserves model accuracy while reducing size and simplifying hardware implementation. Detailed results are presented in Section~\Romannum{5}.

\section{FPGA Implemented result}

This hardware architecture was developed using the Xilinx Vitis High-Level Synthesis (HLS) tool and implemented via the Vivado Design Suite. We realized the five-layer SNN classifier, as depicted in Fig.~\ref{fig: SNN architecture}, on the Xilinx Virtex-7 FPGA VC709 platform. To facilitate efficient hardware deployment, all floating-point weights from the original PyTorch model were quantized to 16-bit fixed-point representations. Furthermore, weight sparsity was introduced through fine-grained pruning.

\subsection{Accumulation Efficiency Analysis}

\begin{table}[htbp]
\centering
\caption{Accumulation Count Ratio Under Different Spatial Sparsity Levels}
\label{tab:accumulation_count_under_different_sparsity}
\setlength{\tabcolsep}{6pt} 
\renewcommand{\arraystretch}{1.2} 

\begin{tabular}{|c|c|c|c|c|}
\hline
\multirow{2}{*}{\textbf{Spatial Sparsity}} & \multicolumn{4}{c|}{\textbf{Accumulation Count Ratio}} \\
\cline{2-5}
 & \textbf{Layer 1} & \textbf{Layer 2} & \textbf{Layer 3} & \textbf{Layer 4} \\
\hline
0\%  & 100.00\% & 100.00\% & 100.00\% & 100.00\% \\
\hline
10\% & 89.70\%  & 89.73\%  & 89.74\%  & 90.00\%  \\
\hline
20\% & 79.83\%  & 79.96\%  & 79.95\%  & 79.96\%  \\
\hline
30\% & 69.80\%  & 69.65\%  & 70.02\%  & 70.13\%  \\
\hline
40\% & 59.87\%  & 59.92\%  & 59.77\%  & 59.85\%  \\
\hline
50\% & 49.85\%  & 49.91\%  & 49.80\%  & 50.10\%  \\
\hline
60\% & 39.74\%  & 39.03\%  & 40.19\%  & 40.14\%  \\
\hline
70\% & 29.80\%  & 30.39\%  & 29.47\%  & 29.97\%  \\
\hline
80\% & 20.01\%  & 19.72\%  & 20.02\%  & 20.07\%  \\
\hline
90\% & 9.89\%   & 9.44\%   & 10.14\%  & 9.79\%   \\
\hline
\end{tabular}
\end{table}

Table~\ref{tab:accumulation_count_under_different_sparsity} illustrates the effectiveness of our sparsity-aware design in reducing accumulation operations, compared to the baseline FINN-based design~\cite{guo2023end}. With the same input, as the spatial sparsity increases from 0\% to 90\%, there is a consistent and significant reduction in the accumulation count ratio across Layer 1 to 4.

This trend demonstrates that leveraging both temporal and spatial sparsity significantly improves accumulation efficiency compared to utilizing only temporal sparsity. Notably, the reduction ratio in accumulation operations closely aligns with the level of spatial sparsity, highlighting the critical role spatial sparsity plays in efficient SNN inference. Layer 5 is excluded from this analysis due to its limited dimensionality, which introduces high variability in accumulation counts and compromises the reliability of the results. Overall, these results underscore the substantial advantages of incorporating sparsity-aware mechanisms into SNN hardware accelerators.

\subsection{Comparison with Prior Works}

\begin{table*}[t]
\centering
\caption{Comparison of Resource Utilization, Performance, and Accuracy Across Various Prior Accelerator Designs.}
\label{tab:design comparison prior}
\begingroup
\renewcommand{\arraystretch}{1.2}
\setlength{\tabcolsep}{4pt}
\resizebox{\textwidth}{!}{%
\begin{tabular}{|c|c|c|c|c|c|c|c|c|c|c|c|c|c|c|c|}
\hline
\textbf{Design} &
\textbf{Model} &
\textbf{Platform} &
\textbf{Q-Bit} &
\textbf{\#LUT} &
\textbf{\#FF} &
\textbf{\#BRAM} &
\textbf{\#DSP} &
\makecell[c]{\textbf{Dynamic}\\\textbf{Power}\\\textbf{(W)}} &
\makecell[c]{\textbf{Static}\\\textbf{Power}\\\textbf{(W)}} &
\makecell[c]{\textbf{Thro.}\\\textbf{(MS/s)}} &
\makecell[c]{\textbf{Energy/S}\\\textbf{(nJ/S)}} &
\makecell[c]{\textbf{Fmax}\\\textbf{(MHz)}} &
\textbf{Accu.} &
\makecell[c]{\textbf{Latency}\\\textbf{($\mu$s)}} &
\makecell[c]{\textbf{FoM}\\\textbf{($\mu$J/S)}} \\
\hline
\cite{emad2021deep} & 4-ANN & ZCU104 & 16 bit & 89512 & 57726 & N/A & 1116 & \textbf{0.254} & {\textbf{0.593}} & 4.78 & {53.14} & {70}  & N/A & N/A & 4756.4 \\
\hline
\cite{wang2024mr} & TFM & AXU3EG & 16 bit & \textbf{20060} & \textbf{11511} & 167 & \textbf{115} & 0.6 & {N/A} & N/A & {N/A} & {N/A} & lower & N/A & N/A\\
\hline
\cite{wang2024mr} & TFM & AXU3EG & 32 bit & 47629 & 36239 & 251 & 261 & 0.6 & {N/A} & 0.704 & {852.27} & {N/A} & \textbf{0.86} & 500 & 40592.9 \\
\hline
\cite{guo2023end} & 5-SNN & Virtex 709 & 16 bit & 74578 & 54289 & \textbf{63.5} & 118 & 1.146 & {1.063} & 11.45 & {100.09} & {\textbf{137}}  & \textbf{0.86} & \textbf{454.85} & 7464.3 \\
\hline
SAOCDS 100 & 5-SNN & Virtex 709 & 16 bit & 82859 & 44906 & 96.5 & 297 & 0.473 & {1.038} & \textbf{23.5} & {\textbf{20.13}} & {\textbf{137}} & \textbf{0.86} & 3246.42 & \textbf{1667.8} \\
\hline
SAOCDS 50 & 5-SNN & Virtex 709 & 16 bit & 84467 & 44815 & 85.5 & 297 & 0.493 & {1.039} & \textbf{23.5} & {20.98} & {\textbf{137}} & 0.85 & 1640.98 & 1772 \\
\hline
SAOCDS 15 & 5-SNN & Virtex 709 & 16 bit & 85671 & 44991 & 76.5 & 297 & 0.552 & {1.040} & \textbf{23.5} & {23.49} & \textbf{137} & 0.85 & 497.94 & 2012.4 \\
\hline
\end{tabular}
}
\endgroup
\end{table*}

\begin{equation}
    \label{eq: FoM}
    FoM=\#LUT*\frac{Dynamic ~ Power}{Throughput}
\end{equation}

In Table \ref{tab:design comparison prior}, we present the implementation results for the RadioML 2016 accelerator. Hardware utilization is characterized by the number of look-up tables (LUTs), flip-flops (FFs), block RAMs (BRAMs), and digital signal processing units (DSPs).

Table \ref{tab:design comparison prior} reports both dynamic and static power to characterize logic activity and the baseline FPGA consumption. Dynamic power is estimated via the tool Vivado Design Suite using SAIF (Switching Activity Interchange Format) files generated from a cycle-accurate simulation of the accelerator design, ensuring accurate calculation based on actual switching activity. Analyses under voltage/temperature corners assume commercial-grade parts, maximum process variation, 25 °C ambient temperature, and standard supply voltages.

Throughput is measured as the number of input I/Q samples processed per second (S/s) at the maximum operating frequency ($F_{max}$), with each dataset frame containing 128 I/Q sample pairs. For fair comparison of power efficiency over designs, we compute energy per sample using the measured dynamic power and throughput. To include hardware resource utilization in the evaluation, we define a Figure of Merit (FoM) (see Equation~\ref{eq: FoM}), which rewards lower LUT usage and dynamic power together with higher throughput. Accordingly, a smaller FoM indicates a more efficient accelerator in overall performance, power, and resource usage.

Among the FPGA resource metrics, we use LUT utilization as the primary indicator of logic consumption. This choice is justified because LUTs (together with associated registers) form the configurable logic fabric in FPGAs — they implement arbitrary combinational and sequential logic, and are the fundamental building blocks for custom logic on the chip. In contrast, FFs largely reflect storage or pipeline registers rather than logic complexity, BRAMs primarily account for storage capacity rather than logic operations, and DSP blocks in our design are used only for auxiliary neuron-state update (e.g., membrane potential decay), not for the core accumulation path. Therefore, LUT count most directly reflects the amount of custom logic and serves as the most representative metric for logic resource consumption in our comparison.

Regarding classification accuracy, note that the average accuracy on the RadioML 2016 dataset tends to be low \cite{o2016radio, jung2022chip, guo2023end} (around 57\%), particularly under low SNR conditions. In \cite{wang2024mr}, the reported accuracy represents the maximum achieved at high SNR. While the results reported in \cite{guo2023end} and our design are obtained at high SNR after quantization and pruning.

For the 4-layer ANN model proposed in \cite{emad2021deep}, 16-bit quantization is applied. However, the hardware utilization, especially of LUTs, FFs, and DSPs, remains relatively high. Although the dedicated use of DSPs enables efficient MAC operations on an FPGA and helps achieve relatively low dynamic power, DSPs are also a scarce and valuable resource. Because the throughput is low, the energy per sample remains high, indicating poor power efficiency. Consequently, while the reported FoM appears favorable, it overlooks the heavy DSP usage.

The two Transformer-based implementations from~\cite{wang2024mr} achieve high simulation accuracy without quantization. However, once 16-bit quantization is applied, their accuracy significantly drops, rendering effective modulation recognition infeasible, despite improved latency. Although these designs exhibit relatively low hardware utilization, their throughput is severely limited due to high computational complexity in Transformer~\cite{fournier2023practical}, especially without quantization and pruning. As a result, their FoM values and energy/sample are less competitive. It is worth mentioning that none of the first three accelerators employed streaming architectures, so their throughputs are relatively low.

The prior streaming architecture from~\cite{guo2023end} adopts a FINN-based streaming approach, achieving relatively high throughput and an improved FoM. Note that in our experiments, the hardware utilization slightly differs from~\cite{guo2023end} because we adopted a longer bit-width for partial products to enhance the resilience against overflow. Additionally, onboard inference accuracy was guaranteed by applying DSPs to implement node-specific decay factors, while accumulation operations remained DSP-free. Thanks to high parallelism, this design also maintains low latency.

Note that in streaming architectures, memory bandwidth rarely becomes a bottleneck. Intermediate activations are passed between layers via on-chip FIFOs, and weights are stored on-chip or synthesized, so off-chip traffic remains minimal. On-chip accesses are modest because each local memory block handles low-rate accesses independently. Consequently, neither off-chip nor on-chip memory requires the high bandwidth typical of systolic-array, so memory bandwidth is generally not reported for streaming architectures.

In our SAOCDS designs at 15\%, 50\% and 100\% uniform density for all layers, the architecture adeptly handles sparsity in both IFMs and kernels. Notably, even at 100\% density (where kernel sparsity is absent) SAOCDS-100 consumes only 41\% of the dynamic power compared to the design in~\cite{guo2023end}. This efficiency stems from reduced data fetching and storage. The streaming nature of the architecture facilitates high throughput and significantly improves the FoM. However, SAOCDS-100 exhibits higher latency than FINN-based streaming architectures, primarily due to its iteration mechanism over non-zero weights. As sparsity increases, latency decreases substantially while maintaining strong FoM performance. At 85\% sparsity, our design achieves comparable latency to~\cite{guo2023end} while consuming only 48.2\% of its power.

In summary, compared to prior AMC research using the RadioML 2016 dataset, our SAOCDS design effectively exploits sparsity in both IFMs and kernels, consistently delivering the highest throughput and achieving low power consumption with only a modest increase in hardware resources. While latency is higher with dense kernels, it scales proportionally with kernel density and declines significantly as sparsity increases. At very high levels of spatial sparsity, SAOCDS achieves very low latency while preserving its highest throughput, competitive classification accuracy, and low power consumption.

\subsection{Detailed Comparison across Different Weight Densities}

\begin{table*}[t]
\centering
\caption{Comparison of Resource Utilization, Performance, and Accuracy between SAOCDS under Various Spatial Sparsity Levels.}
\label{tab:design comparison}
\begingroup
\renewcommand{\arraystretch}{1.2}
\setlength{\tabcolsep}{6pt}
\resizebox{\textwidth}{!}{%
\begin{tabular}{|c|c|c|c|c|c|c|c|c|c|c|c|c|}
\hline
\textbf{Design} & 
\textbf{\#LUT} & 
\textbf{\#FF} & 
\textbf{\#BRAM} & 
\textbf{\#DSP} & 
\makecell[c]{\textbf{Dynamic} \\ \textbf{Power} \\ \textbf{(W)}} &
\makecell[c]{\textbf{Static} \\ \textbf{Power} \\ \textbf{(W)}} &
\makecell[c]{\textbf{Thro.} \\ \textbf{(MS/s)}} &
\makecell[c]{\textbf{Energy/S} \\ \textbf{(nJ/S)}} &
\makecell[c]{\textbf{Fmax} \\ \textbf{(MHz)}} &
\makecell[c]{\textbf{Latency} \\ \textbf{($\mu$s)}} &
\makecell[c]{\textbf{Accuracy} \\ \textbf{(\%)}} &
\makecell[c]{\textbf{FoM} \\ \textbf{($\mu$J/S)}} \\
\hline
\cite{guo2023end} 100 & \textbf{74578} & 54289 & 63.5 & \textbf{118} & 1.146 & {1.063} & 11.45 & {100.09} & {137} & 454.85 & \textbf{100} & 7464.3 \\
\hline
SAOCDS 100 & 82859 & 44906 & 96.5 & 297 & 0.473 & {1.038} & \textbf{23.5} & {20.13} & {137} & 3246.42 & \textbf{100} & 1667.8 \\
\hline
SAOCDS 75  & 83322 & 44902 & 89.5 & 297 & 0.432 & {1.036} & \textbf{23.5} & {18.38} & {137} & 2460.18 & 99.98  & 1531.7 \\
\hline
SAOCDS 50  & 84467 & \textbf{44815} & 85.5 & 297 & 0.493 & {1.039} & \textbf{23.5} & {20.98} & {137} & 1640.98 & 99.51 & 1772.0 \\
\hline
SAOCDS 25  & 83868 & 44901 & 79.0 & 297 & 0.481 & {1.038} & \textbf{23.5} & {20.47} & {137} & 822.10  & 99.22 & 1716.6  \\
\hline
SAOCDS 20  & 83514  & 44915  & 78.5  & 297 & 0.541 & {1.040} & \textbf{23.5} & {23.02} & {137} & 658.90  & 99.17 & 1922.6 \\
\hline
SAOCDS 15  & 85671  & 44991  & 76.5  & 297 & 0.552 & {1.040} & \textbf{23.5} & {23.49} & {137} & 497.94  & 97.64 & 2012.4 \\
\hline
SAOCDS 10  & 84175  & 44945  & \textbf{75.0}  & 297 & 0.473 & {1.037} & \textbf{23.5} & {20.13} & {137} & \textbf{453.14}  & 93.33 & 1694.3 \\
\hline
SAOCDS 5   & 83572  & 45060  & 75.5  & 297 & \textbf{0.361} & {\textbf{1.033}} & \textbf{23.5} & {\textbf{15.36}} & {137} & \textbf{453.14}  & 73.19 & \textbf{1283.8} \\
\hline
SAOCDS 25-20-15-20-25  & 84808  & 44969  & 77.0  & 297 & 0.615 & {1.043} & \textbf{23.5} & {26.17} & {137} & 500.82  & 98.19 & 2219.4 \\
\hline
SAOCDS 20-15-10-15-20  & 84297  & 45011  & 76.5  & 297 & 0.579 & {1.041} & \textbf{23.5} & {24.64} & {137} & \textbf{453.14}  & 95.68 & 2076.9 \\
\hline
\end{tabular}
}
\endgroup
\end{table*}

Comparing performance across different spatial-density levels offers deeper insight into the efficiency of our accelerator. We include the baseline FINN-based design~\cite{guo2023end}, which uses the same SNN architecture and IFM dataflow, for direct comparison. Table \ref{tab:design comparison} summarizes the results: each entry names the accelerator and its layer-wise density. A single value denotes uniform density across all five layers; a sequence (e.g., “25-20-15-20-25”) specifies per-layer weight densities. During training, we apply layer-specific pruning ratios to achieve these target densities, and we ensure that at deployment each layer attains its assigned density.

\subsubsection{Resource Utilization Analysis}

The SAOCDS design shows stable LUT and FF usage across different spatial sparsity levels, while BRAM usage decreases as sparsity increases. This is because the logical resources are synthesized to support a single iteration, and sparsity only affects the number of iterations—not the resource demand per iteration. As fewer non-zero weights need to be stored at higher sparsity, BRAM consumption decreases accordingly.

Moreover, SAOCDS utilizes more DSPs due to its higher degree of parallelism in handling potential decay operations. Specifically, it performs 297 decay operations across all layers compared to 118 in the baseline, confirming proportional scaling of DSP resources in line with increased computational parallelism.

\subsubsection{Power, Latency, and Throughput Analysis}
SAOCDS maintains relatively stable power consumption across most sparsity levels, with two notable exceptions: a sharp drop at 5\% uniform density and an increase with mixed layer-wise densities. This is because sparsity affects only the number of iterations, not per-iteration resource usage. The FPGA’s static power remains nearly constant across all designs (1.033–1.043 W), meaning that variations in dynamic power are the primary reflection of logic complexity and resource usage.

Latency decreases approximately in proportion to the reduction in density, consistent with the behavior of the GOAP-based convolutional design. Since the third convolutional layer has the highest iteration count, its latency dominates the total latency of the streaming accelerator. However, at very high sparsity levels (e.g., 5\%), latency plateaus due to a shift in the bottleneck to the FC layer. Unlike GOAP, the WM method used in the FC layer skips redundant operations to reduce power but cannot lower latency. This results in pipeline stalls, ultimately leading to reduced overall power consumption.

In mixed-density designs, layer-wise workload is better balanced, allowing more logic units to operate simultaneously. Since no blocking occurs, power consumption increases due to higher average utilization, as the convolutional layer remains the bottleneck.

Throughput remains consistent across all sparsity levels, as the streaming architecture ensures fully pipelined parallel execution. Since the critical path is unaffected by sparsity, throughput is stable and higher than~\cite{guo2023end} due to a more efficient implementation of the slowest pipeline stage.

\subsubsection{Accuracy Analysis}

In our classification accuracy comparisons, we evaluate the compressed models against their original PyTorch counterparts, rather than against ground-truth labels.

With joint pruning and quantization during training, moderate sparsity has minimal impact on classification accuracy. Even at 10\% uniform density, the model retains relatively high accuracy, though performance drops significantly at lower densities. Introducing layer-wise density allows higher sparsity in larger layers, helping balance the workload, reduce latency bottlenecks, and preserve accuracy. For instance, compared to the baseline~\cite{guo2023end}, the SAOCDS 25-20-15-20-25 configuration achieves near-optimal latency with less than 2\% accuracy loss and only 53.7\% of the baseline’s power consumption—while maintaining similar resource usage and 2$\times$ throughput—representing the best overall trade-off among the evaluated designs.

\section{Conclusion}

In this article, we propose a sparsity-aware output-channel dataflow streaming inference SNN accelerator optimized for high-throughput and low-power real-time applications, such as AMC in cognitive radio systems. To the best of our knowledge, this is the first fully-pipelined streaming SNN architecture that integrates both temporal and spatial sparsity. The accelerator iterates only over non-zero weights in the kernels and performs accumulation only when the corresponding inputs are non-zero, thereby eliminating unnecessary weight fetching and accumulation operations. By doing so, our accelerator achieves automatic load balancing across PEs and supports high throughput even under constrained hardware resources.

The fixed nature of the kernels during inference allows potential empty and extra iterations to be precomputed, enabling seamless streaming across layers without the router and scheduler overhead typical of traditional systolic array-based accelerators. Implemented on an FPGA, the proposed SAOCDS accelerator achieves a throughput of 23.5 MS/s, outperforming previously reported accelerators on the RadioML 2016 dataset.

Compared to the FINN-based streaming SNN accelerator baseline, the proposed design achieves 2× higher throughput while consuming only 41.3\% of the dynamic power at 100\% kernel density, with only an 11.1\% increase in logic hardware utilization. As kernel sparsity increases, the accelerator maintains high throughput and low dynamic power consumption, and its latency decreases markedly—becoming even lower than the baseline under high spatial sparsity. Furthermore, classification accuracy remains relatively high across a wide range of sparsity levels, demonstrating the robustness and efficiency of the design.

\bibliographystyle{ieeetr}

\bibliography{reference}

\end{document}